\documentclass[prl,a4paper,twocolumn,showpacs,showkeys,superscriptaddress,amsmath,amssymb]{revtex4}
\usepackage{graphicx}
\usepackage{dcolumn}
\usepackage{bm}
\usepackage[dvips]{color} 
\begin{document}
\title{Bose analogs of MIT bag model of hadrons in coherent precession}

\author{S.~Autti}
\affiliation{Low Temperature Laboratory, School of Science and Technology, Aalto University, Finland}


\author{V.B.~Eltsov}
\affiliation{Low Temperature Laboratory, School of Science and Technology, Aalto University, Finland}

\author{G.E.~Volovik}
\affiliation{Low Temperature Laboratory, School of Science and Technology, Aalto University, Finland}
\affiliation{L.D. Landau Institute for Theoretical Physics, Moscow, Russia}

\date{\today}
\begin{abstract}
  Recently it was demonstrated that magnon condensation in the trap exhibits the phenomenon of self-localization \cite{Autti2012}. When the number of magnons in the textural trap increases, they drastically modify the profile of the gap and highly increase its size. The trap  gradually transforms from the initial harmonic one to the box with walls almost  impenetrable for magnons. The resulting texture-free ``cavity'' filled by the magnon  condensate wave function becomes the bosonic analog of the MIT bag, in which
hadron  is seen as a cavity  surrounded by the QCD
vacuum, in which the free quarks are confined in the ground or excited state.  Here we consider the bosonic analog of the MIT bag with quarks on the ground and excited levels.

\end{abstract}

\pacs{67.30.er, 03.70.+k, 05.30.Rt, 11.10.Lm}


\maketitle

In Ref. \cite{Autti2012} it was demonstrated that the Bose-Einstein condensation (BEC) of magnons in the flexible trap exhibits the phenomenon of self-localization with formation of a box. This phenomenon is not unique in nature. Other examples of self-formation of a box-like trapping potential are the electron bubble in liquid helium  and the MIT bag
model of a hadron \cite{Chodos1974}, where the asymptotically free
quarks are confined within a cavity surrounded by the QCD
vacuum. Here we discuss the analogy of magnon BEC with the MIT bag model.

The MIT bag model has been used for construction of different hadrons, including mesons, baryons and even multiquark hadrons, such as tetraquarks \cite{Jaffe1977} and pentaquarks \cite{Strottman1979}. In the MIT bag model, free quarks are forced to move only inside a given spatial region, within which they occupy single-particle orbitals. MIT bag is described by the following energy whose minimization determines the equilibrium radius $R$ of  a given hadron:
 \begin{equation}
E(R)=\sum_a N_a \sqrt{m_a^2c^4 + \frac{\hbar^2c^2x_a^2}{R^2}}  + F(R) ~~,~~ F(R)=B \frac{4\pi R^3}{3} \,.
\label{MITbag}
\end{equation}
Here the first term is the kinetic energy of quarks with masses $m_a$ in the cavity of radius $R$, where parameters $x_a$ are determined by the boundary conditions for fermions on the boundary of the bag. For the fermion in the ground state  in the box, and  in the ultra-relativistic limit of vanishing fermionic masses, $m_a\rightarrow 0$, the parameter $x = 2.04$. The second term is the potential energy, $B$ is the so-called bag constant that reflects the bag pressure. At zero temperature, the bag constant $B$ is the difference in the energy density between the false vacuum inside the bag (the deconfinement phase) and the true QCD vacuum outside  (the confinement phase).
In the non-relativistic limit, ignoring the term which does not depend on $R$, one gets
 \begin{equation}
E(R)=\sum_a N_a  \frac{\hbar^2 x_a^2}{2m_aR^2}  + F(R) \,.
\label{MITbag_Non-Rel}
\end{equation}
The same equation describes the electron bubble in superfluid $^4$He, where $m$ is the electron mass; $x=\pi^2$ for the ground state level; the potential energy $F(R) = (4\pi/3)R^3P +4\pi \sigma R^2$, with $P$ being the external pressure and $\sigma$  the surface tension.
For extension to the multi-electron bubbles in superfluid $^4$He see \cite{SalomaaWilliams1981}.

\begin{figure}[t]
\includegraphics[width=\linewidth]{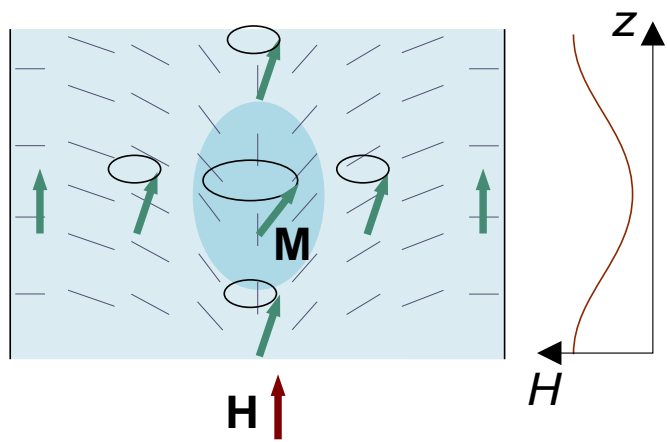}
\caption{Original magneto-textural trap for magnons. The  oscillator frequency $\omega_z$ of the well in the axial direction in \eqref{Potential2} can be regulated by the field in the pinch coil. The frequency $\omega_r$ in the radial direction can be adjusted by applying rotation, since the vortex-free superfluid flow or the array of rectilinear vortex lines created by rotation modifies the flare-out texture
$\hat{\bf l}(r)$. }
 \label{InitialTrap}
\end{figure}

This consideration is applicable for magnon BEC.
The original harmonic trap for magnons is  schematically shown in Fig.~\ref{InitialTrap} \cite{Autti2012}. The confinement potential  $U_\parallel(z)=\gamma H(z)$ in the axial direction is  produced by local perturbation of magnetic field with a small pinch coil.  In the radial direction the well $U_\perp(r)$ is formed by the cylindrically symmetric flare-out texture of the orbital vector ${\bf l}$ (direction of the orbital momentum of Cooper pairs). It comes from  the  spin-orbit interaction energy \cite{BunkovVolovik2010} 
  \begin{equation}
U_\perp(r)|\Psi|^2= \frac{2\Omega_\mathrm{L}^2}{5\omega_\mathrm{L}}\left(1-l(r)\right)  |\Psi|^2 \,,
\label{OrientationalEnergy}
\end{equation}
where $\Psi$ is the wave function of magnon condensate, and  $l=\hat{\bf l}\cdot \hat{\bf H}\equiv \cos \beta_L$ describes the orientation of the orbital momentum $\hat{\bf l}$ with respect to the direction $\hat{\bf H}$ of magnetic field.
On the side wall of the cylindrical container the orbital momentum $\hat{\bf l}$ is normal to the wall,  while in the center it is parallel to the axially oriented applied magnetic field. This produces a minimum of the potential $U_\perp(r)$ on  the cylinder axis.
So the total confinement potential for the $\Psi$-field is
  \begin{equation}
  U({\bf r})= U_\parallel(z) + U_\perp(r)= \omega_\mathrm{L}(z) +
\frac{2\Omega_\mathrm{L}^2}{5\omega_\mathrm{L}}\left(1-l(r)\right) \,.
\label{Potential1}
\end{equation}
 Close to the axis the polar angle $\beta_L$ of
the $\hat{\bf l}$-vector varies linearly with distance $r$ from the axis. As a result, the
potential  $U({\bf r})$ reduces to that of a usual harmonic trap used for the
confinement of dilute Bose gases \cite{PitaevskiiStringari2003}:
 \begin{equation}
U({\bf r})= U(0) + \frac{m_\mathrm{M}}{2} \left( \omega_z^2 z^2 + \omega_r^2 r^2\right)\,,
\label{Potential2}
\end{equation}
where we take into account that the axial trap is also close to harmonic. The harmonicity of the initial trap was checked by measurement of the spectrum of magnons --  standing spin waves in the trap, which was found to have nearly equidistant levels. The equidistant levels of spin precession localised around the minimum of the external magnetic field
have been derived in Ref. \cite{KupkaSkyba2003} using the full set of Leggett equations for spin dynamics of $^3$He-B.

\begin{figure}[t]
\includegraphics[width=\linewidth]{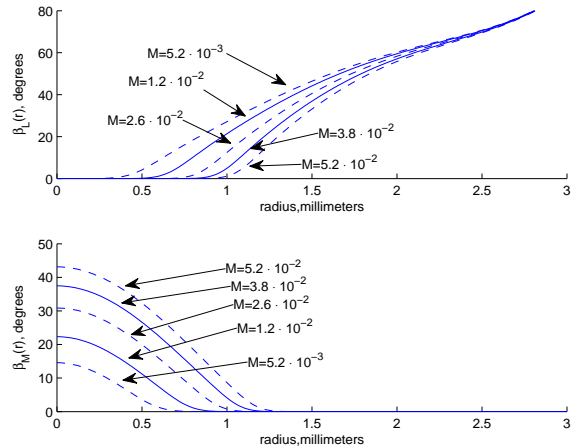}
\caption{Simulation of formation of the multi-magnon bubble with magnon condensate on the ground state level, $n_r=0$, in the cell of radius $R_c=3$ mm in the vessel rotating with angular velocity
$\Omega= 0.8$ rad/s in the vortex-free state for different number of pumped of magnons. ({\it top}): $\beta_L$ is the polar angle of the orbital $\hat{\bf l}$-vector, $\cos \beta_L(r)=\hat{\bf l}({\bf r})\cdot \hat{\bf z}\equiv l(r)$. It plays the role of neutral field, which
provides the trap potential for the charged field $\Psi$ according to \eqref{OrientationalEnergy}.
({\it bottom}): $\beta_M(r)$ is the tipping angle of precessing magnetization. 
It is related to the charged field -- the  wave function of magnon BEC --  as $\Psi \sim \sin (\beta_M/2) \exp(i\omega t + i\alpha)$,
see Ref. \cite{BunkovVolovik2010}.  Here $\omega$ and $\alpha$ are the frequency and the phase of coherent precession, which in the magnon BEC language are the chemical potential and the phase of the Bose condensate correspondingly.    The total transverse magnetization  $M_\perp= \int dV \chi H  \sin \beta_M(r)$ is connected with the number of magnons   ${\cal N}=\int dV |\Psi|^2
\propto \int dV (1-\cos \beta_M(r))$.  In figure, $M$ denotes the normalized magnetization, $M=V^{-1} \int dV \sin \beta_M(r)$. With increasing  $M$ and thus with increasing  ${\cal N}$, the neutral field $\beta_L(r)$ is more effectively repelled from the region of the magnon condensate due to repulsive interaction in \eqref{OrientationalEnergy}. As a result, the harmonic trap gradually transforms to the box ({\it top}), while the wave function of the condensate $\Psi$ gradually transforms to the Bessel function ({\it bottom}).  }
 \label{angles_Ground}
\end{figure}

In the analogy with the MIT bag, the flexible texture of the orbital momentum $\hat{\bf l}$ plays  the role of either the pion field or the non-perturbative gluonic field depending on the microscopic structure of the confinement phase.
The trap is modified by pumped magnons due to spin-orbit interaction in  \eqref{OrientationalEnergy}:
 the condensation of field $\Psi$ in the trap leads to the preferable   orientation of  $\hat{\bf
l}$ parallel to magnetic field, $l=1$, in the region of the trap. In other words the condensate repels the $\hat{\bf l}$-field from the region, where magnons are localized. At large number ${\cal N}$ of magnons in the trap the systems becomes similar to MIT bag with cavity free from the orbital field, which is occupied by magnons. 
So magnons, like quarks,  dig a hole pushing the orbital field away due to the repulsive interaction,
see Fig. \ref{angles_Ground}.
The main difference from the MIT bag model is that magnons are bosons and may macroscopically occupy the same energy state in the trap, forming the  Bose-condensate, while in MIT bag the number of fermions on the same energy level is limited by the Pauli principle. The bosonic bag becomes equivalent to the fermionic bag in the limit of large number of quark flavors, when ${\cal N}Ê\gg 1$ quarks  may occupy the same level. Without loosing generality, we may consider the 2D approximation, i.e the 2D cylindrical trap. In the limit of large ${\cal N}$, the radius $R$ of the cavity filled with ${\cal N}$ magnons occupying the quantum state with  radial number $n_r$ is determined by a balance of two terms in the total energy of the bag \cite{Autti2012}:
 \begin{equation}
E(R,n_r)={\cal N} \epsilon_{n_r}(R)+ F(R)~~,~~ \epsilon_{n_r}(R)=\frac{\hbar^2 \lambda_{n_r+1}^2} {2m_\mathrm{M}R^2} \,.
\label{balance}
\end{equation}
The first term on the rhs is magnon zero-point energy in the cavity. It is the magnon number ${\cal N}$
in the Bose condensate times the energy $\epsilon_{n_r}$ of a single magnon on the radial level $n_r$ in the cylindrical box
with impenetrable walls. In NMR experiments only the
energy levels with zero azimuthal quantum number are excited. They are measured as the shift  of the frequency  of the NMR peak, corresponding to excitation of a magnon,  with respect to the Larmor frequency $\omega_L$:
\begin{equation}
\Delta\omega=\omega-\omega_L=\frac{\epsilon_{n_r}}{\hbar}= \frac{\hbar \lambda_\mathrm{{n_r}+1}^2} {2m_\mathrm{M}R^2}    \,.
\label{R_vs_frequency}
\end{equation}
Here $m_\mathrm{M}$ is  as before the magnon mass, and the parameter $x$ in \eqref{MITbag_Non-Rel} equals the $n_r+1$-th root of the Bessel function, $x=\lambda_\mathrm{n_r+1}$, which corresponds to the proper boundary condition for the magnons populating the radial level $n_r$ in the impenetrable box. The potential energy $F(R)$  in \eqref{balance} corresponds to the pressure exerted to the bag by the field of the orbital texture, which is expelled from the bag. It is the difference in the energy of the orbital field texture with and without the cavity.  

Experimental results for the ground state magnon condensate in \cite{Autti2012} demonstrated that they can be reproduced by the phenomenological equation 
\eqref{balance} if one assumes that there is the scaling law  $F(R)\propto R^k$.  Minimization of the phenomenological equation Eq.~\eqref{balance} with respect to $R$ 
suggests that at large ${\cal N}$ the radius of localization approaches the asymptote:
\begin{equation}
R({\cal N})\sim ~a_r \left({\cal N}/{\cal N}_\mathrm{c}\right)^{1/(k+2)}~~,~~{\cal N} \gg {\cal N}_\mathrm{c}\,,
\label{boxrad}
\end{equation}
where $a_r$ is the harmonic oscillator length in the original radial trap (at ${\cal N} \ll {\cal N}_\mathrm{c}$), ${\cal N}_\mathrm{c}$ is the characteristic number at which the scaling starts,
  In experiments, the  dependence of the transverse magnetization ${\cal M}_{\perp}$ on the frequency shift $\Delta\omega$ is measured. As distinct from  the 
magnon number  ${\cal N}=\int d^2r |\Psi|^2$, where $\Psi$ is the  wave function of magnon condensate, 
 the transverse magnetization density represents the order parameter and is proportional to $\Psi$. The total magnetization is thus  ${\cal M}_{\perp}\propto \int d^2r |\Psi| \propto {\cal N}^{1/2} R$. Since $\Delta\omega  \propto 1/R^2$ according to \eqref{R_vs_frequency}, one obtains ${\cal M}_{\perp}\propto (\Delta\omega)^{-1-k/4}$. The measured transverse magnetization suggests that for large ${\cal N}$ the scaling law is approached with $k\approx 3$. The magnetization estimated in numerical simulations also suggests that under experimental conditions the $k=3$ scaling is the reasonable fit  \cite{Autti2012},  which is just the scaling corresponding to the MIT bag model.
 
In the above approach  the  energy consideration has been used, where the energy potential  is obtained by averaging over fast precession. The numerical simulations of the $Q$-ball, which used the full dynamical equations for spin and order parameter in one-dimensional textural trap, can be found in \cite{Bunkov2005}.

Incidentally, for an atomic condensate in harmonic trap the radius $R$ as a function of number of atoms at large ${\cal N}$ also approaches the scaling  in Eq.~(\ref{boxrad})  with $k=3$ \cite{PitaevskiiStringari2003}. This behavior results from the repulsive inter-particle interactions in the Thomas-Fermi limit.  
However, this similarity in the scaling law $k=3$ both with atomic condensate and with MIT bag model is accidental.
Moreover, the $k=3$ scaling is actually in disagreement with the typical bag models.  If the main contribution to the pressure comes from the bulk, as it happens for the hadron model, then the energy $F(R)$ should be proportional to the volume of the bag, and then for the  two-dimensional radial trap one would expect $F(R) =\pi R^2 P$, i.e. $k=2$.  On the other hand, if the main contribution to the pressure comes from the surface tension, as it happens for electron bubble in liquid helium at $P=0$,  then the energy $F(R)$ should be proportional to the surface area, and for our 2D case this would give $F(R)=2\pi \sigma R$, i.e. $k=1$.  The observed more soft behavior with approximate scaling law  $k\approx 3$ in the 2D case reflects the flexibility of the orbital field, which is inhomogeneous outside the cavity. On general grounds, $F(R)$ depends on several length scales: radius $R$ of the bubble; radius $R_{\rm c}$ of the cylindrical container, where the boundary conditions on the $\hat{\bf l}$-texture are imposed; and the textural healing lengths: magnetic  length $\xi_H$ (the thickness of the layer near the wall of the container
in which the orientation of $\hat{\bf l}$ by magnetic field is restored) and the lengths related to the orientational effects of rotation and vortices on $\hat{\bf l}$.  

\begin{figure}[t]
\includegraphics[width=\linewidth]{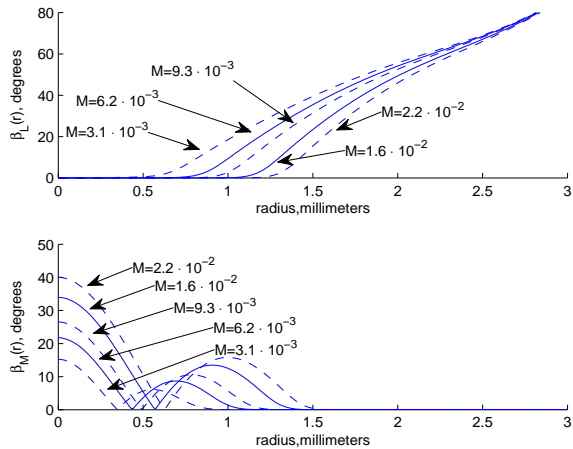}
\caption{The same as in Fig. \ref{angles_Ground} but with the condensate formed at the first excited radial level in the trap, $n_r=1$.  The flat region with $l=1$ develops at large transverse magnetization ({\it top}), indicating the formation of a box in which magnons are condensed on the first excited level ({\it bottom}). The wave function $\Psi$ of the condensate has a node. }
 \label{angles_1stExcited}
\end{figure}

We investigated numerically the scaling law emerging for the condensates formed on the  ground and excited levels for textures in the rotating vessel of two different sizes.
In the MIT bag model the shape of the bag is spherical if all quarks are in the ground state.   When considering excited hadrons in which quarks occupy also the higher energy levels, the non-spherical shapes of the bag must also be considered. In our case this would correspond to the magnon bubble with broken axial symmetry. However, so far in our  NMR experiments we were able to excite and grow BEC condensates only with zero azimuthal quantum numbers, so that the shape of the bag with such condensates is automatically cylindrical. 
Figs. \ref{angles_Ground} and \ref{angles_1stExcited} demonstrate numerical simulations of the formation of the magnon condensates on correspondingly ground state level and on the excited radial level in the original harmonic trap formed by the vortex-free texture in the experimental cell of radius $R_c=3$mm in the cryostat rotating as $\Omega=0.8$ rad/s. At large number of pumped magnons the original harmonic trap is  modified: for all radial quantum numbers $n_r$ the trap gradually transforms to the cylindrical box free from the $\hat{\bf l}$ texture. 

\begin{figure}[t]
\includegraphics[width=\linewidth]{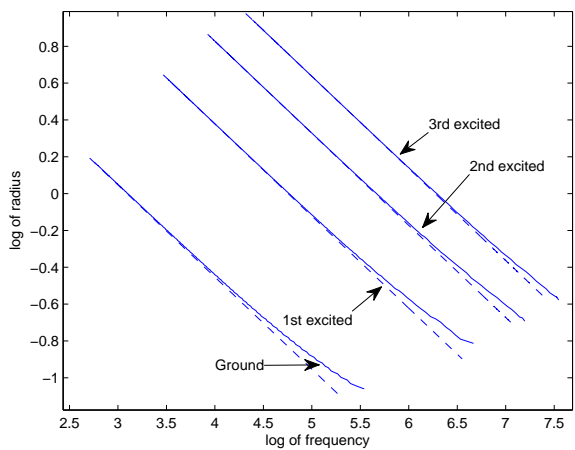}
\caption{Log-Log plot of the radius of the magnon bubble as a function of the frequency shift for the magnon BEC in the ground state  and on three excited levels. Solid lines -- numerical simulations, dashed lines correspond to Eq. \eqref{R_vs_frequency} describing the energy levels in a cylindrical box.
}
 \label{radiusVSfrequency}
\end{figure}

We checked how well  the modified trap represents the cylindrical box with the wall impenetrable for magnons.
For such a box  filled with magnons on $n$-th level the radius as a function of the frequency shift is given by \eqref{R_vs_frequency}.
 Figure \ref{radiusVSfrequency} demonstrates the dependence of $R$ on the frequency shift for different radial quantum numbers $n_r$ of the magnon condensate in the trap both from numerical simulations and 
from the phenomenological equation  \eqref{R_vs_frequency}. One can see that with decreasing frequency the harmonic states of the condensate gradually approach the states in the box which obey Eq. \eqref{R_vs_frequency}.

Both in experiments and in the numerical simulations in  \cite{Autti2012}, all the length scales were of the same order, and thus no really small parameter was available, which could justify the scaling law
with $k\sim 3$.  The true scaling behavior may only appear in some limit cases. One of the limit cases may take place  in the  large vessel rotating with angular velocity $\Omega$ in a vortex-free state in the regime, when $R_c\gg R \gg \xi_v \gg \xi_H$, where $\xi_v$ is the healing length related to counterflow $|{\bf v}_s-{\bf v}_n|=\Omega r$, which orients $\hat{\bf l}$ in the plane. In the large vessel with counterflow the boundary of the vessel are not important, because the main  orientational effect comes from the counterflow, so that within the cavity one has  $l=1$ due to spin-orbit interaction \eqref{OrientationalEnergy} of the $\hat{\bf l}$-field with the condensate, while outside the cavity one has $l=0$ due to counterflow. As a result, the energy $F(R)$ is simply the excess of the orientational energy of  the counterflow in the cavity
$F(R)\propto \pi R^2(\Omega R)^2$, i.e. the scaling law with $k=4$ emerges in the limit of large vessel, or large counteflow. This gives for the radius of the magnon BEC bag   at large ${\cal N}$
the asymptote
\begin{equation}
R \sim {\cal N}^{1/6}\,,
\label{1/6}
\end{equation}

\begin{figure}[t]
\includegraphics[width=\linewidth]{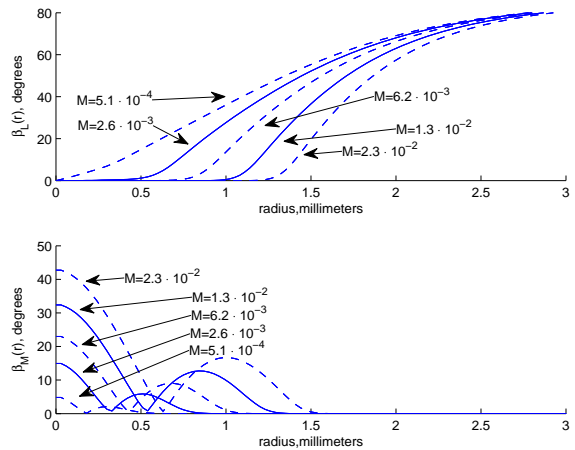}
\caption{The same as in Fig. \ref{angles_1stExcited} for the first excited state, but in the larger vessel
(radius $R_c=30$ mm). On the plot only the central part of texture is shown: for $r> 3$ mm there is  nothing interesting to be seen, since the texture practically reaches the constant value $l=0$ prescribed by the counterflow, while the wavefunctions are equal to zero.}
 \label{LV1stExcited}
\end{figure}

We probed this regime in our simulations using the vortex free texture in the rotating vessel with large radius $R_c=30$ mm.
Figure \ref{LV1stExcited} demonstrates the texture and the condensate density in case of the magnon BEC on the first excited level. Qualitatively the behavior is similar to that in
Fig. \ref{angles_1stExcited}, obtained
for the experimental cell with the radius $R_c=3$ mm. The power law in the dependence $R({\cal N})$ of radius of the magnon BEC bag calculated for the ground state condensate, is
 \begin{equation}
R \sim {\cal N}^{0.178}\,.
\label{0}
\end{equation}
This power is more close to $1/6\approx 0.167$  in \eqref{1/6}, which corresponds to $k=4$, rather than to $0.2$ for $k=3$.

The MIT bag model is used also for description of large (macroscopic) box with the deconfinement
phase inside, which is filled  with thermal quarks and gluons forming the quark-gluon plasma, 
and with the hadronic  phase outside, which contains only massive particles, such as pions. Using macroscopic  MIT box one can describe the confinement-deconfinement phase transition, see e.g.\cite{Fraga2012}.
In our case the analog of such transition occurs, when
the magnon density reaches the value at which the orientational effect of the magnon BEC on the orbital field
exceeds the competing orientational effect of the counterflow in the whole vessel. This happens when
\begin{equation}
\sin\beta_M \sim  \frac{\omega_L}{\Omega_L}   \frac{\Omega R}{v_{\rm pb}}  \,,
\label{condition}
\end{equation}
where $v_{\rm pb}$ is the pair breaking velocity and $\Omega_L$ is the so-called Leggett 
frequency, which characterizes the spin-orbit interaction. When this threshold is reached, the texture (which is an analog of the pion
field) is completely wiped out of the vessel, and the whole
vessel becomes occupied by the magnon BEC --  the counterpart of plasma of free quarks and gluons.
However, the coherent precession in the whole vessel is known to experience Suhl instability
 -- the catastrophic relaxation \cite{Catastropha,Catastroph2,Catastroph3}.
  
In conclusion, the extension of the MIT bag model to bosons allows us to describe the self-localization of magnon Bose-condensate. The magnon bag is the kind of the $Q$-ball \cite{Friedberg1976} formed due to interaction between the magnon condensate described by the field $\Psi$ with conserved charge $Q={\cal N}$,  and the orbital field $\chi(r)=1-l(r)$, which is the analogue of the neutral field \cite{BunkovVolovik2007}.  In the process of self-localization the charged field $\Psi$ modifies locally the neutral field $\chi$ so that the potential well is formed, in which the charge
${\cal N}$ is condensed. Due to the repulsive interaction between these field in $^3$He-B, in the limit of large ${\cal N}$ the cavity is formed, which is void of neutral field and is filled with the charge field. 
As a result one obtains the bosonic analog of the hadron in the MIT bag model: instead of the quarks in the ground state in a box potential, there are magnons which also fill the ground state. But since magnons are bosons, they form the Bose condensate in a box. We also applied the MIT bag model to the analogue of an excited hadron -- the bag with magnons condensed in one of the excited levels in a box.

This work is supported by the Academy of Finland and the EU -- FP7 program (\# 228464 Microkelvin).



\end{document}